\documentclass[conference]{IEEEtran}
\IEEEoverridecommandlockouts
\usepackage{cite}
\usepackage{comment}
\usepackage{amsmath,amssymb,amsfonts}
\usepackage{algorithm}
\usepackage{algorithmic}

\usepackage{booktabs}
\usepackage{graphicx}
\usepackage{textcomp}
\usepackage{xcolor}
\usepackage{float}
\usepackage{multirow}
\usepackage{adjustbox}
\usepackage{cleveref}
\usepackage{subcaption}
\usepackage{flushend}
\usepackage[caption=false]{subfig}
\usepackage{geometry}

\def\BibTeX{{\rm B\kern-.05em{\sc i\kern-.025em b}\kern-.08em
    T\kern-.1667em\lower.7ex\hbox{E}\kern-.125emX}}

\DeclareFontFamily{U}{stix2bb}{}
\DeclareFontShape{U}{stix2bb}{m}{n} {<-> stix2-mathbb}{}

\begin{document}

\title{RADEP: A Resilient Adaptive Defense Framework Against Model Extraction Attacks}
\author{
    \IEEEauthorblockN{
    Amit Chakraborty \IEEEauthorrefmark{3},
    Sayyed Farid Ahamed\IEEEauthorrefmark{1},
    Sandip Roy\IEEEauthorrefmark{1}\IEEEauthorrefmark{4} ,
    Soumya Banerjee\IEEEauthorrefmark{1}\IEEEauthorrefmark{4},
    }
    \IEEEauthorblockN{
    Kevin Choi\IEEEauthorrefmark{2}, 
    Abdul Rahman\IEEEauthorrefmark{2},
    Alison Hu\IEEEauthorrefmark{2},
    Edward Bowen\IEEEauthorrefmark{2},
    Sachin Shetty\IEEEauthorrefmark{1}
    }
    \IEEEauthorblockA{
    \IEEEauthorrefmark{3}Department of CSBS, Asansol Engineering College, Asansol, West Bengal 713305, India
    \\idmeamit@gmail.com}
    \IEEEauthorblockA{
    \IEEEauthorrefmark{1}Center for Secure \& Intelligent Critical Systems, Old Dominion University, Virginia, USA \\
    \IEEEauthorrefmark{4}School of Cybersecurity, Old Dominion University, Virginia, USA
    \\\{saham001, sroy, s1banerj, sshetty\}@odu.edu}
    \IEEEauthorblockA{
    \IEEEauthorrefmark{2}Deloitte \& Touche LLP
    \\ \{kevchoi, abdulrahman, aehu, edbowen\}@deloitte.com}
}

\maketitle
\begingroup\renewcommand\thefootnote{\textsection}
\endgroup

\begin{abstract}
Machine Learning as a Service (MLaaS) enables users to leverage powerful machine learning models through cloud-based APIs, offering scalability and ease of deployment. However, these services are vulnerable to model extraction attacks, where adversaries repeatedly query the application programming interface (API) to reconstruct a functionally similar model, compromising intellectual property and security. Despite various defense strategies being proposed, many suffer from high computational costs, limited adaptability to evolving attack techniques, and a reduction in performance for legitimate users. In this paper, we introduce a \underline{R}esilient \underline{A}daptive \underline{D}efense Framework for Model \underline{E}xtraction Attack \underline{P}rotection (RADEP), a multifaceted defense framework designed to counteract model extraction attacks through a multi-layered security approach. RADEP employs progressive adversarial training to enhance model resilience against extraction attempts. Malicious query detection is achieved through a combination of uncertainty quantification and behavioral pattern analysis, effectively identifying adversarial queries. Furthermore, we develop an adaptive response mechanism that dynamically modifies query outputs based on their suspicion scores, reducing the utility of stolen models. Finally, ownership verification is enforced through embedded watermarking and backdoor triggers, enabling reliable identification of unauthorized model use. Experimental evaluations demonstrate that RADEP significantly reduces extraction success rates while maintaining high detection accuracy with minimal impact on legitimate queries. Extensive experiments show that RADEP effectively defends against model extraction attacks and remains resilient even against adaptive adversaries, making it a reliable security framework for MLaaS models.

\end{abstract}


\section{Introduction}
With the growing popularity of MLaaS, advanced models are now easily accessible via cloud-based Application Program Interface (API)s, bypassing the need for local training but also introducing new vulnerabilities. Recent studies have shown that these models are prone to extraction attacks \cite{juuti2019prada}, where adversaries repeatedly query the victim model using carefully crafted or surrogate inputs to reconstruct a substitute model that mimics its functionality. This unauthorized replication not only jeopardizes intellectual property but can also facilitate further adversarial exploits and compromise user privacy. Machine learning models developed and deployed for various critical infrastructure applications are increasingly targeted by adversarial threats, including ME attack \cite{thakur2023novel}, \cite{das2022securing}. Adversaries may employ both black-box (without internal insight) methods (such as the JBDA-TR \cite{juuti2019prada}, Cloudleak \cite{yu2020cloudleak}, KnockoffNet \cite{orekondy2019knockoff}, Zeroth Order Optimization \cite{chen2017zoo} etc.), and many white-box (transparent or accessible model) techniques to perform these attacks.  Although defenses like adversarial training \cite{jiang2023comprehensive}, model pruning \cite{uchida2017embedding}, and query detection \cite{jiang2023comprehensive} have been proposed, they incur significant computational overhead. Another common challenge with model extraction defenses is the trade-off between security and performance, where strict security measures can degrade the accuracy of the source model \cite{ahamed2024accuracy}, \cite{truong2021datafree}. Moreover, static defenses often fail against adaptive adversaries who modify their strategies to bypass security measures \cite{jiang2023comprehensive}. MLaaS platforms implement strong authentication mechanisms, such as API keys, OAuth, and multi-factor authentication, to restrict model access to authorized users \cite{vangala2022blockchain}. However, an authenticated adversary can still execute model extraction by issuing an excessive number of queries.  
This necessitates a strong research focus on developing a enhanced defense technique to protect MLaaS models while ensuring their integrity and confidentiality.

To address these challenges, we propose RADEP, a multi-layered defense framework that integrates complementary techniques to protect MLaaS models from extraction and privacy attacks. RADEP combines \textit{progressive adversarial training} for resilience against evolving threats, \textit{malicious query detection} using uncertainty and behavioral analysis, and an \textit{adaptive query response} mechanism that perturbs suspicious queries while preserving utility for legitimate users. It also includes \textit{ownership verification} via embedded backdoor triggers and lightweight watermarking, enabling reliable detection of unauthorized model usage.
The main contributions of this paper are as follows:
\begin{itemize}
    \item We propose RADEP, a multifaceted defense framework against model extraction attacks, integrating progressive adversarial training, malicious query detection, adaptive query response, and ownership verification. Through detailed analysis and experiments, we demonstrate how RADEP effectively reduces the success of extraction attacks and limits the utility of stolen models.

    \item We introduce a query detection system that uses uncertainty metrics and behavioral analysis to potentially malicious queries, and a dynamic response mechanism that degrades adversarial outputs.

    \item We ensure malicious query detection time to be less than 0.01 ms and adaptive query response times between 15 ms (MNIST) and 60 ms (ImageNette), with ownership verification completed within 520.5 to 850.5 ms. This low overhead enhances scalability, making RADEP suitable for deployment in resource-constrained MLaaS environments.
    
    \item We conduct extensive experiments to evaluate RADEP’s effectiveness against state-of-the-art model extraction attacks, including \textit{JBDA-TR} \cite{juuti2019prada}, \textit{Cloudleak} \cite{yu2020cloudleak}, and \textit{KnockoffNet} \cite{orekondy2019knockoff}. The results demonstrate RADEP’s superior resilience across different datasets and attack scenarios.

\end{itemize}

The remaining part of this paper is as follows: \Cref{threatModel} outlines the threat model, detailing the attacker's objectives, knowledge, and capabilities. \Cref{overview} describes the proposed RADEP framework. \Cref{results} provides the experimental results along with an analysis and discussion of the findings. Lastly, we conclude our work and discuss a few future research thoughts in \Cref{conclusion}.

\section{Threat Model} \label{threatModel}
In this section, we outline the threat model, detailing the adversary’s objective, knowledge and strategy of the proposed model extraction attack defense framework \cite{jiang2023comprehensive}.
\subsection{Adversary Objective}
Model stealing attacks typically aim to replicate various aspects of a target model, such as its architecture, hyperparameters, or overall functionality \cite{jiang2023comprehensive}. In this paper, we focus specifically on functionality stealing, where the attacker’s goal is to build a substitute model that closely matches the performance of the victim model. To assess the success of such attacks, we use two metrics: (i) test accuracy, which measures how well the substitute model performs on the victim model's test data, and (ii) fidelity, defined as the level of agreement between the outputs of the victim and the extracted model on identical inputs. An ME attacker might be primarily interested in replicating the victim model to avoid ongoing API costs, 
or might use a successful extraction to facilitate further attacks, such as adversarial or membership inference attacks.

\subsection{Adversary Knowledge}
We assume that the adversary has limited access to data and can only interact with the victim model via a black-box (without internal insight) interface \cite{banerjee2024mia}. In this context, ``data-limited" implies that the attacker only has access to a small set of natural samples, while ``black-box access" (without internal insight) means that the attacker can only submit inputs and observe the corresponding outputs from the victim model. Based on the type of outputs received, ME attacks can be categorized into two scenarios: (i) the hard-label scenario, where only the predicted class is returned, and (ii) the soft-label scenario, where the full probability distribution is provided. In our evaluations, we test our defense framework under both of these conditions.

\subsection{Adversary Strategy}
The adversary, limited by a small number of natural samples, overcomes data scarcity by either generating synthetic data or by employing surrogate data to query the victim model. The resulting query-output pairs are then used to train a substitute model. In our evaluation, we consider three advanced attack strategies. First, in the \textit{JBDA-TR} attack, an enhanced version of Jacobian-based dataset augmentation is employed \cite{juuti2019prada}. For each sample in the training set, synthetic samples are iteratively generated using a targeted variant of the Fast Gradient Sign Method (FGSM) \cite{goodfellow2014explaining}, where a random target class is selected in each iteration. The synthetic samples produced are then labeled by the victim model and added to the training data to retrain the substitute model. Second, the \textit{Cloudleak} \cite{yu2020cloudleak} attack similarly relies on synthetic sample generation but differs by using a feature-based adversarial attack to create these samples; the adversary subsequently fine-tunes a pre-trained substitute model with the newly obtained labeled data. Finally, the KnockoffNet attack \cite{orekondy2019knockoff} bypasses synthetic sample generation by querying the victim model with surrogate data from the same or a related distribution and training the substitute model on the responses. These strategies collectively enable the adversary to effectively replicate the victim model’s functionality despite having only limited natural data.

\begin{figure*}[!htb]
    \centering
    \vspace{-15pt}
    \includegraphics[width=0.7\linewidth]{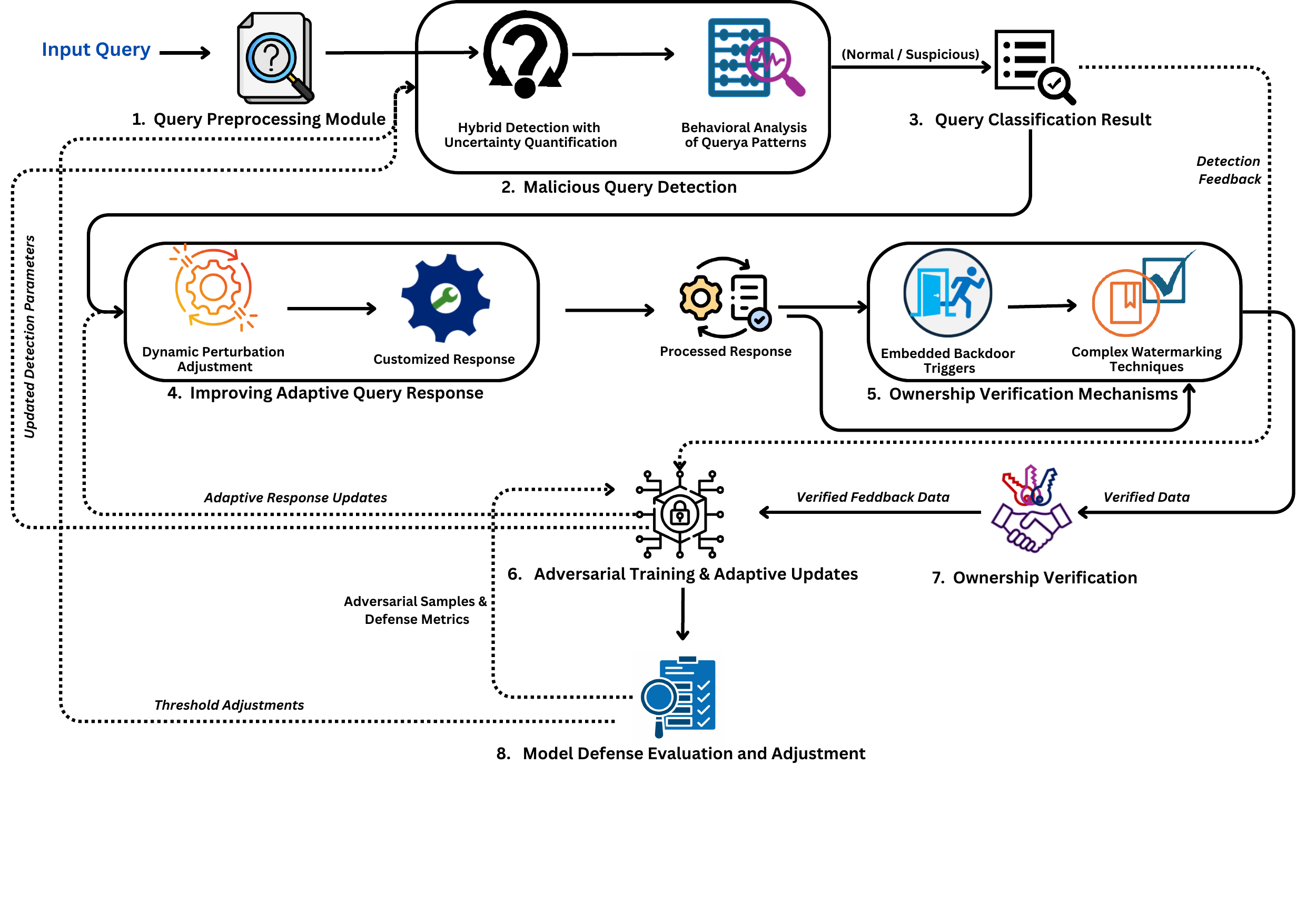}
    \vspace{-40pt}
    \caption{Architecture of the proposed framework RADEP.}
    \label{fig:2}
    \vspace{-10pt}
\end{figure*}

\section{Proposed RADEP} \label{overview}
This section provides a detailed description of each phase of the proposed RADEP framework. Figure \ref{fig:2} illustrates the architecture, highlighting the interconnection of each phase.   

\subsection{Progressive Adversarial Training}
Adversarial training is a key defense mechanism that improves model resilience by exposing the model to diverse adversarial perturbations during training. 
Our approach integrates multiple adversarial techniques, including Fast Gradient Sign Method (FGSM) \cite{goodfellow2014explaining}, Projected Gradient Descent (PGD) \cite{madry2017towards}, and DeepFool \cite{ducoffe2018adversarial}, to generate adversarial examples that challenge the model’s decision boundaries. For example, FGSM perturbs an input \(x\) in the direction of the gradient of the loss function \(J(\theta, x, y)\) as follows:
\begin{equation}
x_{\text{adv}} = x + \epsilon \cdot \text{sign}(\nabla_x J(\theta, x, y))
\end{equation}
where \(\epsilon\) is a small perturbation factor, \(\theta\) represents the model parameters, and \(J(\theta, x, y)\) is the loss function. PGD extends this concept by applying iterative updates to refine the adversarial example, while DeepFool minimizes the \(l_2\)-norm to produce more precise perturbations \cite{juuti2019prada}. In addition to these methods, our approach incorporates adaptive training updates that periodically generate and integrate new adversarial examples into the training process. This continuous update mechanism allows the model to adapt to emerging threats without requiring complete retraining, thereby enhancing its resilience against increasingly sophisticated extraction attacks.

\subsection{Malicious Query Detection Mechanisms}
The proposed malicious query detection mechanism combines uncertainty quantification and behavioral analysis to flag malicious queries while preserving normal user experience. In the uncertainty component, a composite score is computed as:

\vspace{-3mm}

\begin{equation}
U(x) = \alpha_1(1 - P_{\text{max}}(x)) + \alpha_2 H(x) + \alpha_3(1 - M(x)) + \alpha_4 \sigma(x)
\end{equation}

\vspace{-1mm}

where \(P_{\text{max}}(x)\) refers to the maximum softmax probability \cite{kariyappa2020defending}), and the entropy $H(x)$ is computed as:  

\vspace{-5mm}

\begin{equation}
H(x) = -\sum_{i=1}^{K} P(y=i \mid x, \theta) \log P(y=i \mid x, \theta)
\end{equation}
where \(M(x)\) is the margin between the top two predictions, and \(\sigma(x)\) is the Bayesian-based uncertainty \cite{maddox2019simple} (e.g., via Monte Carlo Dropout). The weights \(\alpha_i\) are calibrated using validation data, and queries with \(U(x) > \tau\) are flagged as suspicious.

On the other side, the behavioral analysis segments queries into fixed intervals that measure query frequency and variance. Abnormal query rates or unusually low variance, identified using statistical methods such as Kullback-Leibler (KL) divergence \cite{kullback1951information}, indicate potential extraction attempts. Combined, these two approaches create a resilient detection system with high accuracy and minimal false positives.

\subsection{Adaptive Query Response}
Adaptive query response adjusts output perturbations according to the evaluated suspicion level of each query, as outlined in Algorithm \ref{alg:adaptive_query_response}. In our approach, each query is assigned a suspicion score $S(q_i)$ (as shown in line 2 of Algorithm \ref{alg:adaptive_query_response}), computed through different uncertainty metrics such as softmax probability $P_{\text{max}}(q)$, predicted entropy $H(q)$, margin of top two predictions $M(q)$, and Bayesian-based uncertainty $\sigma(q)$.




Based on this score, the system dynamically adjusts the perturbation level applied to the query response. Higher suspicion results in stronger perturbations through techniques such as label flipping and additive noise \cite{zhang2021label}, while highly suspicious queries undergo adaptive label scaling that redistributes output probabilities to obscure the model’s behavior. The perturbed response is calculated (Algorithm \ref{alg:adaptive_query_response} line 9), ensuring that responses are customized for different levels of query suspicion. RADEP customizes responses by comparing incoming queries to previously flagged ones, degrading outputs for suspicious patterns. Periodic threshold recalibration ensures adaptability to evolving attacks with minimal impact on legitimate users.

\begin{algorithm}[H]
\caption{Adaptive Query Response}
\label{alg:adaptive_query_response}
\begin{algorithmic}[1]
\REQUIRE Set of queries \(q_1, \ldots, q_N\)
\ENSURE Perturbed responses for each query 

\FOR{each query \(q_i\)}
    \STATE \textit{Compute:} \(S(q_i) = \alpha_1(1 - P_{\text{max}}) + \alpha_2 H(q_i) + \alpha_3 (1 - M(q_i)) + \alpha_4 \sigma(q_i)\) \textcolor{blue}{$\vartriangleright$ \textit{suspicion score}}
    \STATE \(\epsilon = \begin{cases} 
    \epsilon_{\text{low}} & \text{if } S(q_i) \le \tau_1 \\
    \epsilon_{\text{medium}} & \text{if } \tau_1 < S(q_i) \le \tau_2 \\
    \epsilon_{\text{high}} & \text{if } S(q_i) > \tau_2
    \end{cases}\)
    \STATE \(P_{\text{pert}}(y \mid q_i) = P(y \mid q_i) + \delta(S(q_i)), \quad \delta(S(q_i)) \sim \mathcal{N}(0, \epsilon)\) \textcolor{blue}{$\vartriangleright$ \textit{Perturbed responses}}
\ENDFOR
\RETURN Perturbed responses $P_{\text{pert}}(y \mid q_i)$.
\end{algorithmic}
\end{algorithm}

\subsection{Ownership Verification Mechanisms}
To confirm model ownership, we embed distinctive signatures into the model through both backdoor triggers \cite{szyller2021dawn} and complex watermarking techniques \cite{uchida2017embedding}. In the backdoor trigger approach, specific trigger queries are crafted to produce distinct, predetermined responses that serve as a digital signature. The model is trained so that for each trigger query \( q_{\text{trigger}}^i \), it produces a designated output \( y_{\text{trigger}}^i \), confirming that if a suspected model returns these outputs, its origin can be verified. These triggers are carefully isolated from the normal input distribution to remain resilient against modifications like pruning or fine-tuning \cite{uchida2017embedding}.

Complementing this, complex watermarking techniques are used to subtly modify the model’s standard outputs, embedding a distributed signature that persists across typical queries. For example, the output probability distribution for a watermark query \( q_w \) is adjusted as
\begin{equation}
P_{\text{watermark}}(y \mid q_w) = P(y \mid q_w) + \epsilon(y, q_w)
\end{equation}
where \( \epsilon(y, q_w) \) is a small, query-dependent perturbation that forms the watermark signature. This combined strategy of backdoor triggers and watermarking creates a dual-layered verification mechanism, providing resilient evidence of ownership even after the model undergoes adversarial modifications.

\subsection{Regular Evaluation and Adjustment}
We implement an automated attack simulation that periodically evaluates the defense system. Various adversarial attacks, such as FGSM \cite{goodfellow2014explaining}, PGD \cite{madry2017towards}, and Zeroth Order Optimization \cite{chen2017zoo}, are simulated to generate adversarial examples. The system logs the attack success rate by comparing misclassified examples with clean data, and monitors the false positive rate to minimize impact on legitimate queries. Successful adversarial examples are added to the training set, and detection thresholds are adjusted if the attack success rate exceeds a set tolerance. This feedback loop continuously refines adversarial training, query detection, and response strategies, maintaining high performance while ensuring resilience against extraction attacks.

\section{Experimental Evaluation}\label{results}
In this section, we first describe our experimental setup, then analyze the impact of adversarial training on model extraction attacks. Next, we compare our detection method with Out of Distribution (OOD) detection \cite{kariyappa2020defending} and evaluate the overall effectiveness of RADEP, using Dynamic Adversarial Watermarking of Neural networks (DAWN) \cite{szyller2021dawn}, deceptive perturbation \cite{lee2019defending}, adaptive misinformation \cite{kariyappa2020defending}, and AMAO \cite{jiang2023comprehensive} as baselines.

\subsection{Experimental Setup}
\subsubsection{Datasets and Victim Models}
We consider LeNet-5 for MNIST, AlexNet for FMNIST, ResNet18 for CIFAR-10, and ResNet34 for ImageNette. These datasets collectively offer varying levels of image complexity and dataset sizes, enabling a comprehensive evaluation of RADEP's effectiveness under diverse experimental settings.

\subsubsection{Attack Configuration}
To strengthen the attack setting and evaluate our defense against a more capable adversary, we configure the substitute model to mirror the victim model's architecture. This design ensures that if RADEP proves effective under such stringent conditions, its resilience would be even greater when the adversary lacks this knowledge.

In \textit{JBDA-TR}\cite{juuti2019prada} and \textit{Cloudleak}\cite{yu2020cloudleak}, the substitute model begins training with 100, 100, 1,000, and 1,000 samples for MNIST, F-MNIST, CIFAR-10, and ImageNette, respectively. The corresponding query budgets for these datasets are 10,000, 10,000, 100,000, and 100,000. JBDA-TR involves six iterative augmentation rounds, with the substitute model being trained for 20 epochs after each round. Alternatively, \textit{Cloudleak} employs a pre-trained substitute model that is fine-tuned for 20 epochs to enhance attack performance.

For \textit{KnockoffNet}~\cite{orekondy2019knockoff}, surrogate datasets are leveraged for training the substitute models: FashionMNIST for MNIST, MNIST for F-MNIST, CIFAR-100 for CIFAR-10, and ImageNet for ImageNette. The assigned query budgets are 60,000, 60,000, 50,000, and 13,000 for the respective datasets. Each substitute model is trained for a total of 50 epochs to improve extraction accuracy.

Our primary goal is to assess the robustness of RADEP against a range of model extraction attacks rather than performing a comparative analysis among the attacks themselves. Although the query budgets vary across different attack strategies due to their unique characteristics, this variation does not affect the fairness of our defense evaluation.

\subsubsection{Metrics}
To evaluate RADEP's performance, we consider test accuracy, fidelity, and watermark accuracy as primary metrics, focusing on both defense effectiveness and model robustness. Additionally, we analyze computational overhead to ensure practical applicability. Since fidelity closely aligns with test accuracy trends, we exclude its detailed results for brevity.

\subsubsection{Comparison with Existing Defenses}
We evaluate RADEP against five state-of-the-art defense strategies to comprehensively measure its effectiveness in mitigating model extraction attacks. PRADA~\cite{juuti2019prada} identifies malicious queries by analyzing Gaussian distance distribution deviations, based on the observation that synthetic queries often cluster around seed samples. OOD Detection~\cite{kariyappa2020defending} classifies adversarial inputs as out-of-distribution using tailored detection methods. DAWN~\cite{szyller2021dawn} introduces a hashing mechanism that perturbs predicted labels, limiting extraction performance while embedding backdoors for ownership verification. Deceptive Perturbation~\cite{lee2019defending} alters probability vectors to mislead attackers without affecting the original classification. AMAO~\cite{jiang2023comprehensive} presents an end-to-end defense framework that protects the model throughout its lifecycle, from training to deployment. Adaptive Misinformation~\cite{kariyappa2020defending} degrades stolen model accuracy by generating incorrect outputs through a reverse model trained with reverse cross-entropy loss.

While these defenses mainly concentrate on query detection or response alteration, RADEP advances beyond them by integrating adversarial training with a hybrid query detection framework. Furthermore, it enhances ownership verification using dual-layer techniques, combining watermarking and backdoor triggers to ensure both robust defense against extraction and reliable tracing of unauthorized model usage.

\begin{table}[t]
    \centering
    \renewcommand{\arraystretch}{1.1}  
    \setlength{\tabcolsep}{4pt}        
    \small                             
    \caption{The performance of the substitute model under the defense of adversarial training, where the most effecgive results are highlighted.} \label{table:substitute model under the defense of adversarial training}
    \resizebox{\linewidth}{!}{  
        \begin{tabular}{c c c | c c}
            \toprule
            \textbf{Attack} & \textbf{Scenario} & \textbf{Dataset} & \textbf{Std. train} & \textbf{Adv. train} \\ 
            \hline \hline
            \multirow{8}{*}{\textbf{\shortstack{\textit{JBDA-TR}}}}  
                & \multirow{4}{*}{Hard}  
                    & MNIST       & 91.23  & \textcolor{blue}{88.93}  \\ \cline{3-5}
                &             & F-MNIST    & 79.33  & \textcolor{blue}{77.87}  \\ \cline{3-5}
                &             & CIFAR-10   & 42.80  & \textcolor{blue}{42.05}  \\ \cline{3-5}
                &             & ImageNette & 51.27  & \textcolor{blue}{47.22}  \\ \cline{2-5}
                & \multirow{4}{*}{Soft}  
                    & MNIST       & 95.58  & \textcolor{blue}{91.89}  \\ \cline{3-5}
                &             & F-MNIST    & \textcolor{blue}{81.44}  & 82.51  \\ \cline{3-5}
                &             & CIFAR-10   & 43.97  & \textcolor{blue}{43.14}  \\ \cline{3-5}
                &             & ImageNette & 55.76  & \textcolor{blue}{54.86}  \\ \hline
                
        \end{tabular}
    }
\end{table}


\subsection{Evaluations on Adversarial Training}
We perform experimental evaluations against \textit{JBDA-TR}, \textit{KnockoffNet}, and \textit{Cloudleak} attacks, with the corresponding results summarized in Table~\ref{table:substitute model under the defense of adversarial training}. The extraction models trained adversarially using RADEP consistently achieve lower accuracy compared to standard models, demonstrating the effectiveness of RADEP’s defense mechanisms in both hard-label and soft-label attack settings. In the hard-label scenario, where adversary information is limited, RADEP significantly restricts model extraction. Even in the soft-label setting, RADEP effectively hinders the adversary’s ability to capture decision boundary information. These findings confirm that RADEP’s progressive adversarial training and adaptive strategies increase the adversary’s query requirements, elevate computational costs, and enhance overall model robustness against extraction attacks.

\begin{table}[t]
    \centering
    \renewcommand{\arraystretch}{1.5}
    \caption{Accuracy and F1-Score comparisons demonstrate that RADEP, highlighted, outperforms OOD detection \cite{kariyappa2020defending} and AMAO \cite{jiang2023comprehensive}.}
    \label{table:Accuracy and F1-Score comparisons} 
    \begin{scriptsize}
    \begin{tabular}{c c | c c}
        \toprule
        \textbf{Dataset} & 
        \textbf{Attack} & 
        \shortstack{\textbf{Accuracy (\%)} \\ \cite{kariyappa2020defending} / \cite{jiang2023comprehensive} / RADEP} & 
        \shortstack{\textbf{F1-Score} \\ \cite{kariyappa2020defending} / \cite{jiang2023comprehensive} / RADEP} \\
        \hline \hline
        \multirow{3}{*}{MNIST}  
            & \textit{JBDA-TR}      & 76.15 / 87.85 / \textcolor{blue}{90.57}  & 0.68 / 0.87 / \textcolor{blue}{0.91}  \\ \cline{2-4}
            & \textit{KnockoffNet}  & 86.20 / 93.75 / \textcolor{blue}{95.86}  & 0.82 / 0.93 / \textcolor{blue}{0.96}  \\ \cline{2-4}
            & \textit{Cloudleak}    & 72.23 / 84.99 / \textcolor{blue}{87.31}  & 0.61 / 0.84 / \textcolor{blue}{0.89}  \\ \hline
        \multirow{3}{*}{F-MNIST}  
            & \textit{JBDA-TR}      & 73.05 / 81.20 / \textcolor{blue}{83.73}  & 0.66 / 0.81 / \textcolor{blue}{0.85}  \\ \cline{2-4}
            & \textit{KnockoffNet}  & 65.91 / 69.89 / \textcolor{blue}{72.54}  & 0.53 / 0.67 / \textcolor{blue}{0.73}  \\ \cline{2-4}
            & \textit{Cloudleak}    & 66.33 / 74.37 / \textcolor{blue}{76.82}  & 0.54 / 0.73 / \textcolor{blue}{0.78}  \\ \hline
        \multirow{3}{*}{CIFAR-10}  
            & \textit{JBDA-TR}      & 73.41 / 78.99 / \textcolor{blue}{81.90}  & 0.67 / 0.76 / \textcolor{blue}{0.82}  \\ \cline{2-4}
            & \textit{KnockoffNet}  & 84.55 / 86.23 / \textcolor{blue}{88.85}  & 0.83 / 0.87 / \textcolor{blue}{0.91}  \\ \cline{2-4}
            & \textit{Cloudleak}    & 77.04 / 77.33 / \textcolor{blue}{80.28}  & 0.73 / 0.76 / \textcolor{blue}{0.80}  \\ \hline
        \multirow{3}{*}{ImageNette}  
            & \textit{JBDA-TR}      & 77.11 / 80.29 / \textcolor{blue}{83.43}  & 0.80 / 0.82 / \textcolor{blue}{0.86}  \\ \cline{2-4}
            & \textit{KnockoffNet}  & 72.58 / 81.13 / \textcolor{blue}{84.57}  & 0.76 / 0.81 / \textcolor{blue}{0.85}  \\ \cline{2-4}
            & \textit{Cloudleak}    & 77.16 / 83.04 / \textcolor{blue}{86.09}  & 0.80 / 0.84 / \textcolor{blue}{0.88}  \\ \hline
    \end{tabular}
    \end{scriptsize}
\end{table}


\subsection{Evaluation on Malicious Query Detection}
RADEP demonstrates superior performance in detecting adversarial queries compared to \textit{AMAO}\cite{jiang2023comprehensive} and \textit{OOD detection}\cite{kariyappa2020defending} approaches. It utilizes a hybrid detection framework that integrates uncertainty quantification with behavioral analysis, enhancing the robustness of its defense. As shown in Table~\ref{table:Accuracy and F1-Score comparisons}, RADEP substantially improves both accuracy and F1-score in detecting malicious queries. For instance, against \textit{KnockoffNet} attacks on MNIST, RADEP attains 95.86\% accuracy and an F1-score of 0.96, outperforming \textit{AMAO} and \textit{OOD detection}. Moreover, RADEP captures temporal patterns in query behavior, including frequency and variance, which helps detect advanced attacks like \textit{Cloudleak}. On MNIST, RADEP achieves 87.31\% detection accuracy, exceeding the performance of both \textit{AMAO} and \textit{OOD detection}. These results highlight RADEP’s capability to efficiently handle adaptive attacks. Additionally, RADEP incorporates an adaptive response mechanism that dynamically alters detection thresholds and response strategies based on the suspicion level of queries, significantly outperforming static defenses and consistently enhancing performance across diverse attacks and datasets.

\begin{table}[!htb]
    \centering
    \renewcommand{\arraystretch}{1.5}
    \caption{The test accuracy of the substitute model under the defense of RADEP and the baseline defenses \cite{kariyappa2020defending}, \cite{lee2019defending}, \cite{szyller2021dawn}, where RADEP outperforms the baselines is highlighted.}
    \label{table:test accuracy of the substitute model under the defense of RADEP} 
    \begin{scriptsize}
    \begin{tabular}{c c c | c c c}
        \toprule
        \textbf{Attack} & 
        \textbf{Scenario} & 
        \textbf{Dataset} & 
        \shortstack{\textbf{No} \\ \textbf{Defense}} & 
        \shortstack{\textbf{Baseline} \\ \textbf{Defense}} & 
        \textbf{RADEP} \\
        \hline \hline

        \multirow{8}{*}{\textbf{\textit{JBDA-TR}}}  
            & \multirow{4}{*}{Hard Label}  
                & MNIST       & 91.23  & 87.30  & \textcolor{blue}{65.17}  \\ \cline{3-6}
            &             & F-MNIST    & 79.33  & 74.21  & \textcolor{blue}{63.52}  \\ \cline{3-6}
            &             & CIFAR-10   & 42.80  & 35.35  & \textcolor{blue}{30.89}  \\ \cline{3-6}
            &             & ImageNette & 51.27  & 47.88  & \textcolor{blue}{43.90}  \\ \cline{2-6}
            & \multirow{4}{*}{Soft Label}  
                & MNIST       & 95.58  & 91.80  & \textcolor{blue}{78.94}  \\ \cline{3-6}
            &             & F-MNIST    & 81.44  & 75.48  & \textcolor{blue}{59.78}  \\ \cline{3-6}
            &             & CIFAR-10   & 43.97  & 40.97  & \textcolor{blue}{34.53}  \\ \cline{3-6}
            &             & ImageNette & 55.76  & 50.70  & \textcolor{blue}{46.35}  \\ \hline

        \multirow{8}{*}{\textbf{\textit{KnockoffNet}}}  
            & \multirow{4}{*}{Hard Label}  
                & MNIST       & 89.57  & 70.44  & \textcolor{blue}{65.51}  \\ \cline{3-6}
            &             & F-MNIST    & 40.38  & 34.95  & \textcolor{blue}{30.76}  \\ \cline{3-6}
            &             & CIFAR-10   & 69.37  & 63.48  & \textcolor{blue}{47.07}  \\ \cline{3-6}
            &             & ImageNette & 55.90  & 50.88  & \textcolor{blue}{42.54}  \\ \cline{2-6}
            & \multirow{4}{*}{Soft Label}  
                & MNIST       & 91.72  & 80.56  & \textcolor{blue}{77.92}  \\ \cline{3-6}
            &             & F-MNIST    & 42.10  & 37.81  & \textcolor{blue}{32.18}  \\ \cline{3-6}
            &             & CIFAR-10   & 73.02  & 71.35  & \textcolor{blue}{68.59}  \\ \cline{3-6}
            &             & ImageNette & 68.18  & 61.45  & \textcolor{blue}{55.33}  \\ \hline

        \multirow{8}{*}{\textbf{\textit{Cloudleak}}}  
            & \multirow{4}{*}{Hard Label}  
                & MNIST       & 83.72  & 73.14  & \textcolor{blue}{65.51}  \\ \cline{3-6}
            &             & F-MNIST    & 76.07  & 67.82  & \textcolor{blue}{61.45}  \\ \cline{3-6}
            &             & CIFAR-10   & 78.15  & 67.59  & \textcolor{blue}{58.86}  \\ \cline{3-6}
            &             & ImageNette & 86.64  & 73.60  & \textcolor{blue}{66.12}  \\ \cline{2-6}
            & \multirow{4}{*}{Soft Label}  
                & MNIST       & 86.36  & 75.93  & \textcolor{blue}{71.10}  \\ \cline{3-6}
            &             & F-MNIST    & 78.26  & 71.33  & \textcolor{blue}{62.88}  \\ \cline{3-6}
            &             & CIFAR-10   & 80.04  & 71.09  & \textcolor{blue}{65.23}  \\ \cline{3-6}
            &             & ImageNette & 88.19  & 78.10  & \textcolor{blue}{69.97}  \\ \hline
    \end{tabular}
    \end{scriptsize}
\end{table}

\subsection{Overall Evaluations on RADEP from End to End}
In this section, we comprehensively evaluate RADEP against \textit{JBDA-TR}, \textit{KnockoffNet}, and \textit{Cloudleak} attacks across multiple datasets.

\subsubsection{Effectiveness of RADEP}
Experimental results indicate that substitute models extracted from adversarially trained victims under RADEP consistently attain lower test accuracy than those from standard trained models. For instance, under the JBDA-TR (Hard Label) attack on MNIST, the accuracy of the substitute model reduces to 65.17\%, while under Cloudleak on CIFAR-10, it declines to 58.86\%. These findings validate the effectiveness of RADEP’s progressive adversarial training with periodic adaptive updates in defending against iterative attacks. As illustrated in Table~\ref{table:test accuracy of the substitute model under the defense of RADAP}, RADEP outperforms existing defenses, such as DAWN~\cite{kariyappa2020defending} for hard-label scenarios and Deceptive Perturbation~\cite{lee2019defending} and Adaptive Misinformation~\cite{szyller2021dawn} for soft-label scenarios, demonstrating significantly reduced attack success rates across the considered model extraction attacks.

Moreover, RADEP’s hybrid query detection combined with dynamic response strategies further minimizes the impact of adversarial queries. For example, under the \textit{KnockoffNet} (Soft Label) attack on ImageNette, the substitute model’s accuracy drops to 55.33\%, while under \textit{Cloudleak} on F-MNIST, it falls to 61.45\%. This adaptive perturbation mechanism disrupts the attack process by making query-based extraction substantially more difficult for adversaries.

\begin{table}[t]
    \centering
    \renewcommand{\arraystretch}{1.0}  
    \setlength{\tabcolsep}{2pt}        
    \scriptsize                        
    \caption{Computational Overhead of RADEP.} \label{table:overhead_RADEP}
    \resizebox{0.8\linewidth}{!}{      
        \begin{tabular}{c | c  c}
            \hline
            \textbf{Dataset} & \textbf{Phase} & \textbf{Overhead} \\ 
            \hline \hline
            \multirow{4}{*}{MNIST} 
                & Adversarial training      & 8.50 (min)  \\
                & Malicious query detection & $<0.01$ (ms) \\
                & Adaptive query response   & 15.00 (ms)  \\
                & Ownership verification    & 520.50 (ms) \\ 
            \hline
            \multirow{4}{*}{F-MNIST} 
                & Adversarial training      & 14.20 (min) \\
                & Malicious query detection & $<0.01$ (ms)  \\
                & Adaptive query response   & 18.90 (ms)  \\
                & Ownership verification    & 550.80 (ms) \\ 
            \hline
            \multirow{4}{*}{CIFAR-10} 
                & Adversarial training      & 85.00 (min) \\
                & Malicious query detection & $<0.01$ (ms)  \\
                & Adaptive query response   & 32.80 (ms)  \\
                & Ownership verification    & 710.40 (ms) \\ 
            \hline
            \multirow{4}{*}{ImageNette} 
                & Adversarial training      & 200.00 (min)\\
                & Malicious query detection & $<0.01$ (ms)  \\
                & Adaptive query response   & 60.00 (ms)  \\
                & Ownership verification    & 850.50 (ms) \\ 
            \hline
        \end{tabular}
    }
\end{table}

Furthermore, RADEP’s ownership verification mechanisms offer an additional layer of protection by degrading the accuracy of extracted models. Specifically, under the \textit{JBDA-TR} (Soft Label) attack on CIFAR-10, the substitute model's accuracy further declines to 34.53\%, while for the \textit{KnockoffNet} (Hard Label) attack on F-MNIST, it drops to 30.76\%. These evaluations collectively demonstrate RADEP’s capability to significantly weaken model extraction attempts while preserving data privacy.



\subsubsection{Computational Overhead of RADEP}
Table~\ref{table:overhead_RADEP} summarizes the experimental results obtained using a high-performance system with an Intel Xeon processor and NVIDIA A100 GPUs, running Ubuntu 20.04 LTS with TensorFlow and PyTorch utilizing CUDA acceleration. Performance metrics were recorded using Python's time module and framework profilers, verifying RADEP’s stability and efficiency in real-time scenarios.

RADEP achieves malicious query detection within less than 0.01ms and adaptive query response times ranging between 15ms and 60ms across various datasets. Ownership verification is completed within 520.5ms to 850.5ms. Since adversarial training is performed offline during model development and ownership verification is invoked only when required, the runtime overhead for each query remains limited to detection and response, ensuring practicality for deployment.

\section{Conclusion}\label{conclusion}
In this paper, we presented RADEP, a multifaceted defense framework for MLaaS that combines progressive adversarial training, malicious query detection, adaptive response mechanisms, and ownership verification to counter model extraction and privacy attacks. By leveraging a multi-layered approach, RADEP reduces attack success rates and degrades adversarial responses while minimally impacting legitimate queries. Future work will focus on reducing latency, exploring advanced uncertainty metrics, and enhancing resilience in distributed settings.


\end{document}